\documentclass[aip,pof]{revtex4-1}
\usepackage{graphicx}

\draft 

\begin{document}


\title{Transient convective instabilities in directional solidification} 



\author{E. Meca}

\author{L. Ram\'{\i}rez-Piscina}
\affiliation{Departament de F\'{\i}sica Aplicada, Universitat Polit\`ecnica de Catalunya, Doctor Mara\~n\'on 44, E-08028 Barcelona, Spain}


\date{\today}

\begin{abstract}
We study the convective instability of the melt during the initial transient in a directional solidification experiment in a vertical configuration. We obtain analytically the dispersion relation, and perform an additional asymptotic expansion for large solutal Rayleigh number that permits a simpler analytical analysis and a better numerical behavior. We find a transient instability, i.e. a regime in which the system destabilizes during the transient whereas the final unperturbed steady state is stable. This could be relevant to growth mode predictions in solidification.

\end{abstract}


\maketitle 


\section{Introduction}
\label{sec-intro}

When a melt solidifies the advancing solid-liquid interface can develop very complex patterns, which induce inhomogeneities in the solid phase. The resulting microstructure is largely responsible for the final properties of the solid.\cite{boettinger00} A significative number of methods employed in Materials Science involve solidification of a melt, and hence prediction of the detailed dynamics of the solidification front turns out to be of capital importance.
Most solidification processes of applied interest can be modelized by the directional solidification configuration. In it, a mixture or alloy is pulled at a controlled velocity inside a thermal gradient. This setup permits to control to a certain degree the growth velocity of the solid phase, which becomes a control parameter of the problem.\cite{Caroli_Book} Even ignoring convection in the melt (or supressing it in experiments) the phenomenology obtained with this setup is very rich, with growth of cells and dendrites (with the presence of sidebranching), but also with less usual structures such as seaweeds or doublons, and non steady behaviors such as traveling waves, banding, etc. One interesting point is that these processes can be history dependent. For instance, in dendritic growth, solutions with different wavelengths are possible for a given growth condition,\cite{warren90}  and the final one is dynamically selected during the initial transient states of the solidification process.\cite{WarrenPRE1993,losert98PRL} The fact is that after setting the final value of the pulling velocity, there is a transient in which a layer of solute is built up ahead of the advancing front, during which the instability can occur.\cite{losert98PNASa,losert98PNASb} In the current understanding of the problem this transient determines the result of the selection process.\cite{WarrenPRE1993,losert98PRL}  Indeed, calculations performed by using the instability of the steady state are known to yield wrong results on the selected wavelength when comparing to experiments.\cite{somboonsuk84,trivedi84}


While thermal and concentration gradients can induce convection in the melt, this effect has received little attention in time-dependent solidification problems, but it has nevertheless been studied by using different approximations in some solidification setups\cite{hwang96,huang97,emms94,hwang04}.

Convection is unavoidable in the case of lateral heating, i.e. when the thermal gradient is placed horizontally. On the contrary we will center here on a vertical configuration, in which the solidifying sample is immersed in a vertical thermal gradient and is slowly extracted from below. 
In this case thermal gradient is stabilizing, but accumulation of solute ahead the solidification front, depending on its relative density, can induce an instability of the  Rayleigh-Benard type. For this later configuration there have been numerous works studying the convective instability of the steady state. Critical parameters and wavelength of the instability were found by Hurle {\it et al.}\cite{HurlePhF1983} A weakly nonlinear analysis by Jenkins\cite{jenkins85b} derived an amplitude equation that supported hexagonal solutions. Riley and Davis\cite{riley89} derived an asymptotic scaling law for the critical wavenumber in the same system. Also, a weakly non-linear analysis was performed resulting in a Sivashinsky equation.  
Relatively complete studies of the problem of convection in
the melt was undertaken by Impey {\it et al.}\cite{impey91} for steady solutions and LeMarec {\it et al.}\cite{lemarec97} for time-dependent states, finding chaotic dynamics. LeMarec also undertook similar simulations in three dimensions,\cite{lemarec96} focusing in the prediction of actual macrosegregation patterns in the final crystal. Effects of the solidification transient on the convective instability were identified by the experiments of Jamgotchian {\it et al.}\cite{jamgotchian01} who obtained segregation patterns induced by thermosolutal convection during a directional solidification experiment.  Here the size of the convective cells did not agree with the steady-state wavelength for the hydrodynamic instability, but was found to be consistent with estimations obtained taking into account the transient dynamics.

In this paper, we address the question of the first convective instability of the melt during the initial transient of a solidification experiment in the vertical configuration. To this end, it is necessary to employ a time dependent solution of the solidification problem, i.e. the free boundary dynamics of a planar front with the appropriate boundary conditions at the interface and the diffusion equation for the solute in the melt. We will use the approximate solution by Warren and Langer,\cite{WarrenPRE1993} which is known to be remarkably good when compared to the exact solution.\cite{CaroliJCG1993} The analytical structure of such solution adapts well for the stability analysis of the flow, which will be performed semi-analytically as a function of time during all the transient.
 In doing so, time is treated as a parameter, thus performing a quasi-static or adiabatic approximation\cite{emms94,WarrenPRE1993}. 
 Furthermore, an asymptotic expansion for large Rayleigh number will permit to gain physical insight on the obtained solutions. One interesting possibility will be the appearance of a transient instability, i.e. an instability occurring during the transient stage but which is not present in the steady state. Due to the fact that any present flow will induce a deformation of the solidification front, such transient instability is also expected to affect the final selected growth mode. We will find a regime in which such instability appears.

\section{The model}
\label{sec-model}

\begin{figure}
\includegraphics[width=\textwidth]{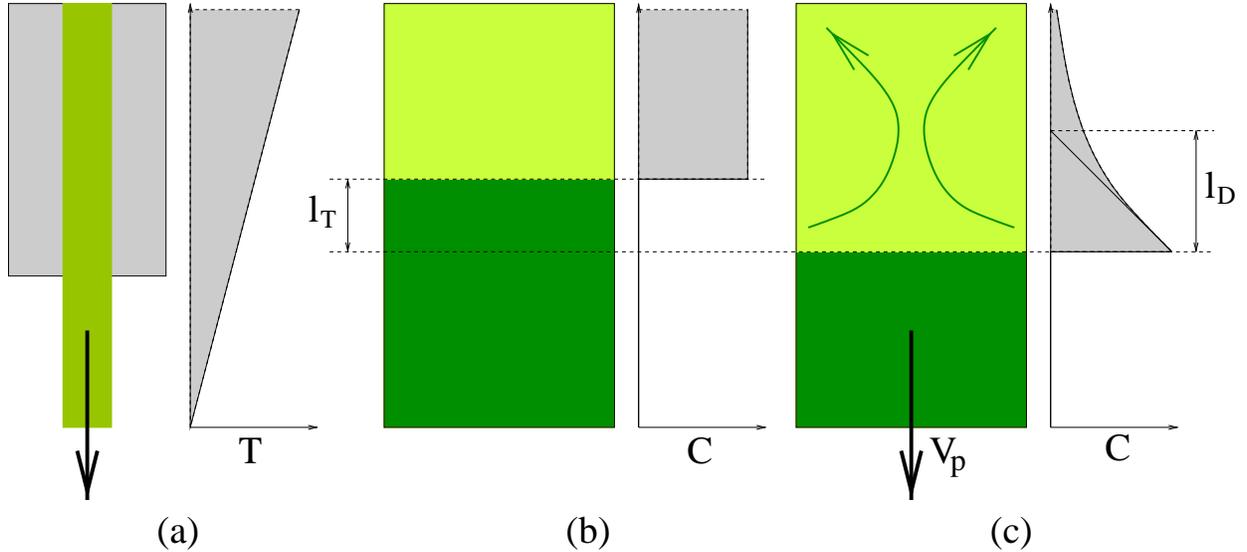}
\caption{(a) Setup of the system. The sample is slowly pulled from a furnace,
  represented as a thermal gradient. (b) Equilibrium configuration and its
  corresponding solute profile, for $K<1$ (c) Steady state configuration and
  concentration profile. The possibility of convection in the melt is also sketched.}
\label{fig:setup_solid}
\end{figure}

In the vertical configuration of a directional solidification experiment we place a sample with a mixture or alloy into a vertical constant thermal gradient as shown in Fig. \ref{fig:setup_solid}a (it is assumed to be time independent owing to the large thermal conductivity. This is the {\it frozen temperature approximation}\cite{LangerRMP1980,Davis_book}). The sample is initially at rest with a constant solute concentration $C_{\infty}$ in the molten phase (Fig. \ref{fig:setup_solid}b). The interface is planar and its initial position is given by the equilibrium temperature at that concentration following the phase diagram of the mixture\cite{Caroli_Book}. The thermal gradient is stabilizing and consequently the melt is quiescent. At time $t=0$ we start to pull from the solid at
a constant speed $v_p$ and, if the interface is stable, a solutal boundary
layer is built ahead of the interface in the steady state (Fig. \ref{fig:setup_solid}c). Note the recoil of the solidification front, due to the change in the interface concentration during the process, and the fact that this means a change in the coexistence temperature.

\subsection{Basic equations}

Mathematical description of this system will consist in the Navier-Stokes equations for the fluid flow, with the appropriate boundary conditions, in the the Boussinesq approximation,\cite{Chandra}
i.e. we will assume the density as a constant mean value except in the buoyancy term. Furthermore we will consider buoyancy effects coming from density changes due to solute concentration only. Effect of thermal buoyancy is stabilizing in the present configuration and would only be of importance at small velocities and solute concentrations.\cite{Davis_handbook} For the solidification process we will use the so-called minimal model of directional solidification,\cite{Caroli_Book} which consists in the diffusion equation for the concentration, supplemented with the boundary conditions (solute conservation and local equilibrium) at the (free moving) interface. All equations will be written exclusively in the liquid phase,
assuming that neither flow or diffusion are possible in the solid phase.

The change in density will depend on concentration through a linear approximation:
\begin{equation}
\rho=\rho_{0}\left(1+\alpha\left(C-C_{ref}\right)\right),
\end{equation}
where $\rho_0$ is a reference density taken at the value of the reference
concentration $C_{ref}$, and $\alpha$ is the solute expansion coefficient, which will be assumed to be negative, corresponding to a buoyant solute.

With the previous approximations taken into account, the governing equations
for the velocity field $\mathbf{v}$ and the solute concentration in the bulk $C$
will take the following form, written in the gradient frame:
\begin{eqnarray}
\partial{}_t \mathbf{v} - v_p \, \partial{}_z \mathbf{v} +
\left(\mathbf{v}\cdot\nabla\right)\mathbf{v} & = &
\mathbf{g}\left(1+\alpha \left(C-C_{ref}\right)\right)- \nabla \pi +\nu \nabla{}^2
\mathbf{v},\label{eq:NS-v}\\
\nabla \mathbf{v} & = & 0,\\
\partial{}_t C - v_p \, \partial{}_z C & = & D
\nabla{}^2 C,
\end{eqnarray}
where $v_p$ is the value of the pulling speed, $\mathbf{g}$ is the acceleration of
gravity, $D$ is the solute diffusion coefficient in the liquid phase, $\nu$ is
the kinematic viscosity and $\pi$ is the pressure reduced by the reference
density. Axis $z$ is placed in the direction of the thermal gradient.

The previous equations are nondimensionalized by using the following
scalings (tildes denote non-dimensional parameters):
\begin{eqnarray}
\begin{array}{lll} {\displaystyle \mathbf{r}}&=&{\displaystyle
   l_D\tilde{\mathbf{r}}},\\ \\
C&=&C_{\infty}\tilde{C}, \end{array} & 
\begin{array}{lll}{\displaystyle t}&=&{\displaystyle \tau_{D}
   \tilde{t}},\\ \\
{\displaystyle \pi}&=&{\displaystyle v_{p}^2\tilde{\pi}}. \end{array}
& \begin{array}{lll} {\displaystyle \mathbf{v}}&=&{\displaystyle
    v_{p}\tilde{\mathbf{v}}},\\ \\ & & \end{array}
\end{eqnarray}
Here $C_{\infty}$ is the initial concentration of the melt. We also use
$l_D=D/v_p$ and $\tau{}_D=D/v_{p}^2$ as the diffusion length and time, respectively. 
Dropping tildes and assuming that we have eliminated the constant terms by
a suitable redefinition of the pressure, taking twice the curl of Eq. \ref{eq:NS-v} and after some manipulation, the following equations are obtained:
\begin{eqnarray}
\partial{}_t \nabla{}^2 \mathbf{v} -  \partial{}_z \nabla{}^2 \mathbf{v} 
&+&
\nabla{}^2\left(\left(\mathbf{v}\cdot\nabla\right)\mathbf{v}\right)-\nabla
\mathrm{N}(\mathbf{v})\,\, = \,\,Sc \nabla{}^2 \nabla{}^2 \mathbf{v} \nonumber\\ 
&+&
Ra_S\frac{K}{1-K}\left(\left(\partial{}_{x}^2C+\partial{}_{y}^2C\right)\mathbf{e_z}-
\partial{}_z \partial{}_y C \mathbf{e_y}-
\partial{}_z \partial{}_x C \mathbf{e_x}\right), \label{eq:NS_DS_rotrot}\\
\partial{}_t C - \partial{}_z C & = & \nabla{}^2 C, \label{eq:diff_DS}
\end{eqnarray}
where
\begin{equation}
\mathrm{N}(\mathbf{v})=(\partial_i v_j)(\partial_j v_i)
\end{equation}
(summation over both indices). The z component of Eq. \ref{eq:NS_DS_rotrot} can be rewritten as:
\begin{eqnarray}
\partial{}_t \nabla{}^2 v_z -  \partial{}_z \nabla{}^2 v_z +
\nabla{}^2\left(\left(\mathbf{v}\cdot\nabla\right)v_z\right)-\partial_z
\mathrm{N}(\mathbf{v}) &=& Sc \nabla{}^2 \nabla{}^2
v_z \label{eq:NS_DS_rotrot_z}\\ 
&+& \nonumber
Ra_S\frac{K}{1-K}\left(\partial{}_{x}^2C+\partial{}_{y}^2C\right).
\end{eqnarray}
Rayleigh number $Ra_S$ and Schmidt number $Sc$ are defined as:\footnote{This definition for 
$Ra_S$ is not the definition used in Ref. \cite{CaroliJPhys1985},
  where $R_s$ is defined as $\frac{Ra_S}{Sc}$, with our convention}
\begin{eqnarray}
 Ra_S&=&\frac{g D \alpha C_{\infty}(K-1)}{K
    v_{p}^3}\\
 Sc&=&\frac{\nu}{D}.
\end{eqnarray}
Here we have introduced the segregation constant $K$. Assuming a dilute alloy the phase diagram is taken as linear, and the concentrations of the coexistent phases at the interface are related through the segregation constant, i.e. $C_{solid}=K C$ at the interface, which we assume to be $K<1$. Note also that $Ra_S$ is positive since we
have a buoyant solute ($\alpha<0$).
Pressure can be recovered from the solution of Eq. \ref{eq:NS_DS_rotrot}
by using the following equation:
\begin{equation}
\nabla{}^{2}\pi=Ra_S\frac{K}{1-K}\partial{}_{z}C+\mathrm{N}(\mathbf{v}). \label{eq:Pressure_DS}
\end{equation}

\subsection{Boundary conditions}
Boundary conditions have to be established for both velocity and concentration fields. 
The velocity field has no-slip boundary conditions at the solidifying interface. Furthermore its normal velocity is zero in the frame moving with the sample. If velocity (in the gradient frame) is redefined subtracting the pulling velocity, it can be seen that all velocity components and their first spatial derivatives do vanish at the interface: 
\begin{eqnarray}
\mathbf{v}(x,y,\zeta(t))&=&0, \label{eq:v-bc}\\
\left.\partial_i v_j \right|_{z=\zeta(t)} &=&0, ~\forall ~ i,j=1\dots 3,\label{eq:dv-bc}
\end{eqnarray}
being $\zeta(t)$ the $z$ coordinate of the planar interface at time $t$. Furthermore, 
since we neglect density changes during the phase transition, the redefined velocity will also vanish at infinity, that is
\begin{equation}
\mathbf{v}=0\,\,\,at\,\,\,z=\infty. \label{eq:infty_DS}
\end{equation}
If pressure has to be recovered from Eq. \ref{eq:Pressure_DS}, a condition
for the normal derivative of the pressure at the interface can be easily found by taking the $z$ component of Eq. \ref{eq:NS-v}.

For the concentration field there are two boundary conditions at the
interface. First of all, we have the solute conservation equation, which in dimensional variables is written as
\begin{equation}
D \partial_z C=\left(K-1\right)C\dot\zeta(t).
\label{eq:sconv_dim}
\end{equation}
The first term stands for the diffusion flux and the term in the right accounts for the solute expelled by the front due to the different equilibrium concentrations of both phases. $\dot\zeta(t)$ is the interface velocity. We have implicitly assumed that
there is no solute diffusivity in the solid.
The other boundary condition is the {\it Gibbs-Thompson} equation, which corresponds to the hypothesis of local equilibrium at the interface and for a planar front reads\cite{Caroli_Book}
\begin{equation}
T_I = T_M + m_L C, \label{eq:gt_dim}
\end{equation}
where $T_I$ is the interface temperature, $T_M$ is the melting temperature of the pure solvent, and
$m_L$ is the (negative) coexistence liquidus slope in the linearized equilibrium phase diagram of the alloy. 
This last equation will permit to relate the concentration to the interface position
in the gradient frame.
We take the origin $z=0$ as the interface position of the steady
state. Due to solute conservation the value of concentration on the solid side of the
interface is $C_{\infty}$ in the steady state, and hence $C_{\infty}/K$ will be the corresponding interface concentration in the liquid side. The steady interface temperature will be then $T_M + m_L {C_{\infty}}/{K}$, and the temperature field can be written as
\begin{equation}
T(z) = T_M + m_L \frac{C_{\infty}}{K} + G_L z,
\label{eq:gradient_temp}
\end{equation}
where $G_L$ is the value of the thermal gradient.

We have finally the two boundary conditions for the concentration. The first is the solute conservation equation, Eq. \ref{eq:sconv_dim}, which written in dimensionless variables reads
\begin{equation}
\partial_z C=\left(K-1\right)C\dot\zeta(t) \label{eq:sconv_ndim}
\end{equation}
The second is the local equilibrium condition at the interface (Eq. \ref{eq:gt_dim}) which, after substituting the temperature from Eq. \ref{eq:gradient_temp}, writing it at the interface ($z=\zeta{}(t)$) and non-dimensionalizing takes the following form:
\begin{equation}
\left(1-\tau\right) \zeta(t) =\frac{1}{K-1}(KC-1) . \label{eq:gt_ndim_z}
\end{equation}
The parameter $\tau$ is defined as:
\begin{equation}
\tau=1-\frac{K D G_L}{\left(K-1\right) v_p m_L C_{\infty}}=1-\frac{l_{D}}{l_{T}},
\end{equation}
where the thermal length $l_T$ is defined as
\begin{equation}
l_{T}=\frac{\left(K-1\right) m_L C_{\infty}}{K G_L}.
\end{equation}
$l_T$ is the scale of the thermal gradient, and in our present setup is also the distance between the interface position at equilibrium (with $v_p=0$) and the (planar) interface position in steady state with $v_p \neq 0$
(see Fig. \ref{fig:setup_solid}).

\subsection{Time dependent solidification problem}

We will consider a linear perturbation of the flow during the initial transient of the directional solidification experiment. The base solution will be then a quiescent fluid, but with time-dependent concentration field, corresponding to the building up of the solute layer ahead of the planar solidification front. The final state of this transient (i.e. for $t \rightarrow \infty$) is the steady state given by
\begin{equation}
C_{st}(z)=1+\frac{1-K}{K}\,e^{-z},~~z \geq 0,
\end{equation}
with the interface placed at $z =0$. An exact solution for the transient without flow can be obtained from the numerical resolution of an integro-differential equation,\cite{CaroliJCG1993} but for our present purposes it is more convenient to use the approximation due to Warren and Langer,\cite{WarrenPRE1993} which is known to be remarkably accurate,\cite{CaroliJCG1993} and is more convenient for the stability calculation. In this approximation we assume the ansatz that the time-dependent concentration has an analytical form similar to that of the steady state. We then write the concentration profile in the following form:
\begin{equation}
C_{WL} (z,t)=1+\left(C_{WL} (\zeta (t),t)-1\right)e^{-\frac{z-\zeta (t)}{l(t)}},
~~z \geq \zeta (t),\label{eq:ansatz_wl}
\end{equation}
where $C_{WL} (\zeta (t),t)$ is the equilibrium concentration corresponding to the local temperature at the interface, $\zeta (t)$ is the interface position, and $l(t)$ is the thickness of the solutal boundary layer ahead of the interface.
By using this ansatz and the
boundary conditions (Eqs. \ref{eq:sconv_ndim}-\ref{eq:gt_ndim_z}), integrating the diffusion equation for the concentration (Eq. \ref{eq:diff_DS}) from $z=\zeta(t)$ to $z=\infty$, and after some manipulation, one obtains two differential equations for the functions $\zeta(t)$ and $l(t)$:\cite{WarrenPRE1993}
\begin{eqnarray}
1+\dot{\zeta}(t)&=&\frac{1}{l(t)}\frac{1-(1-\tau)\zeta(t)}{(1-\tau)(K-1)\zeta (t)+1}\label{eq:WL1}\\
\dot{l}(t)&=&\frac{(1-\tau)(K\zeta (t)+l(t))}{l(t)\left(1+(1-\tau)(K-1)\zeta(t)\right)}- \frac{(1-\tau)l(t)}{1-(1-\tau)\zeta (t)}\label{eq:WL2}
\end{eqnarray}
Initial conditions for these equations are $\zeta(0)=l_T/l_D=(1-\tau)^{-1}$ and $l(0)=0$. 
Eqs. \ref{eq:WL1}-\ref{eq:WL2} are what we will call the Warren-Langer equations.\cite{WarrenPRE1993}
They can be seen as the lowest order result of an expansion in moments of the diffusion equation, and quantitatively it is a very good approximation\cite{CaroliJCG1993}. As we will use this solution, our calculation, being linear in a small perturbation of the flow, will still be zeroth order in that moment expansion. By numerically solving Eqs. \ref{eq:WL1}-\ref{eq:WL2} and introducing the results into Eq. \ref{eq:ansatz_wl}, we have an analytical expression for the unperturbed concentration profile during the transient that will allow for the instability calculation of flow in the melt. This is performed in the next section.

\section{Transient Flow  Instability\label{TransInstFlow}}

\subsection{Perturbative calculation}

Explicitely the perturbations for both the velocity and the diffusive field are introduced as
\begin{eqnarray}
v_z (x,z,t)&=& \epsilon \, v_{1,z} \left(z-\zeta (t)\right)\mathrm{cos}(ax)e^{\sigma t}\\
C(x,z,t)&=&C_{0}(z,t)+\epsilon \, C_1 \left(z-\zeta
  (t)\right)\mathrm{cos}(ax)e^{\sigma t}.
\end{eqnarray}
When writing the previous equations, an exponential growth for the perturbations has been selected, with a growth rate $\sigma$. This assumption is natural, in that we will not go beyond linear perturbation theory.
We also see that the perturbation has a wavelength $a$ in the direction parallel to the interface. Furthermore, we expect the convective instability to be confined to the region near to the front, i.e. where solute gradient is larger, and hence functions $v_{1,z}$ and $C_{0}$ should vanish at infinity.

It should be noted that an important approximation has been performed when specifying the form of the perturbations. This is the adiabatic or quasi-static approximation\cite{WarrenPRE1993,emms94}, and it has been used in many different applications\cite{mullins63,onuki85,CaroliJCG1993,mozos96}. This approximation consists in taking time as a parameter, and hence take the values of $\sigma$, $\zeta(t)$ and $l(t)$ as constants when solving the problem. The validity of the approximation is conditioned to the separation of the timescales associated with the (fast) growth of the perturbations and the (slow) evolution of the flat interface.

Now we introduce the previous equations into Eqs.
\ref{eq:NS_DS_rotrot_z} and \ref{eq:diff_DS} and, after linearization we
obtain, for the linear order in $\epsilon$:
\begin{eqnarray}
\left((1+\dot{\zeta})\frac{d}{dz}+Sc\left(\frac{d^2}{dz^2}-a^2\right)-\sigma\right)\left(\frac{d^2}{dz^2}-a^2\right)v_{1,z}&=&Ra_S
\frac{K}{1-K}a^2 C_1 \label{eq:linear_for_v1_brut}\\
\left((1+\dot{\zeta})\frac{d}{dz}+\frac{d^2}{dz^2}-a^2-\sigma\right)C_1&=&v_{1,z}\partial
_z C_0 \label{eq:linear_for_C1}
\end{eqnarray}

If we denote the differential operators on the left hand side of the previous
equations by $\mathcal{M}$ for Eq. \ref{eq:linear_for_v1_brut} and by $\mathcal{N}$ for Eq.
\ref{eq:linear_for_C1} we can eliminate $C_1$ by applying
$\mathcal{N}$ over Eq. \ref{eq:linear_for_v1_brut}, and we obtain
\begin{equation}
\mathcal{N}\,\mathcal{M}\,v_{1,z}=a^2 \frac{K}{1-K} Ra_S \, v_{1,z}\partial
_z  C_0
\end{equation}

From $v_{1,z}$ and Eq. \ref{eq:linear_for_v1_brut} we can recover $C_1$.
Now, to obtain a closed solution of the previous equation, we substitute $C_0$ by the Warren-Langer approximation Eq. \ref{eq:ansatz_wl}. Also we take $\zeta(t)$ as a known function, computable by integration of Eqs. \ref{eq:WL1}-\ref{eq:WL2}. Then the previous equation takes the form
\begin{equation}
\mathcal{N}\,\mathcal{M}\,v_{1,z}=-\frac{1}{l} a^2 Ra_S \left(1-(1-\tau)\zeta 
\right)e^{-\frac{\xi}{l}} v_{1,z},
\end{equation}
where we assume that operators and functions depend on $\xi=z-\zeta(t)$. The operator $\mathcal{N\,M}$ can be factorized in first order terms, so that
the previous equation is written as
\begin{equation}
\prod_{i=0}^{5}\left(\frac{d}{d\xi}+\tilde{r}_i\right)v_{1,z}=
-\frac{a^2 Ra_S}{l\,Sc} \left(1-(1-\tau)\zeta 
\right)e^{-\frac{\xi}{l}} v_{1,z} \label{eq:linear_for_v1_interm} ,
\end{equation}
in which the $\tilde{r} _i $ take the following values (where in each case even (odd) indexes correspond to the plus (minus) sign choice):
\begin{eqnarray}
\tilde{r}_{0,1} &=&\frac{1}{2}\left((1+\dot{\zeta})\pm\sqrt{(1+\dot{\zeta})^2 +4(\sigma+a^2)}\right) , \nonumber\\
\tilde{r}_{2,3} &=&\frac{1}{2Sc}\left((1+\dot{\zeta})\pm\sqrt{(1+\dot{\zeta})^2 +4Sc(\sigma+Sc\,a^2)}\right) , \label{eq:def_rtilde}\\
\tilde{r}_{4,5} &=&\pm a . \nonumber
\end{eqnarray}
If we make the following changes:
\begin{eqnarray}
r_i&=&1-l\tilde{r}_i,\label{eq:def_rsintilde}\\
s&=&-\frac{l^5 a^2 Ra_S}{Sc} \left(1-(1-\tau)\zeta
\right)e^{-\frac{\xi}{l}},
\end{eqnarray}
and apply them successively to Eq. \ref{eq:linear_for_v1_interm}, the following equation is obtained for $v_{1,z}$:
\begin{equation}
\left[\prod_{i=0}^{5}\left(s\frac{d}{ds}+r_i-1\right)-s\right]v_{1,z}(s)=0 , \label{eq:linear_for_v1}
\end{equation}
which is the Generalized Hypergeometric Equation.\cite{Luke}
Solutions for that equation can be written with the aid of the generalized hypergeometric function, which is defined by the following series:
\begin{equation}
_{p}F_{q}\left(\alpha _1,\, ..,\alpha _p;\rho _1,\,..,\rho _q;s\right)=
\sum_{n=0}^{\infty}\frac{\prod_{j=1}^p \Gamma(\alpha _j +n)\prod_{k=1}^q
  \Gamma(\rho _k)}{\prod_{j=1}^q \Gamma(\alpha _j)\prod_{k=1}^q \Gamma(\rho _k
  +n)}\frac{s^n}{n!}\label{eq:SeriesHyperDef} .
\end{equation}
In our case, Eq. \ref{eq:linear_for_v1}
has 6 linearly independent solutions, which can be
written as
\begin{equation}
U_h(s)=s^{1-r_h} \vphantom{}_0 F _5 (;1+r_q-r_h^*;s) \label{eq:solutions}, ~ h=0,1,\dots,5.
\end{equation}
For convenience we use a notation in which $( \dots + r_q +\dots )$ stands for a set of arguments with $q=0,1,\dots,5$. The star, e.g. $(\dots +r_q-r_h^*)$, indicates that the case $q=h$ is omitted, then counting as 5 arguments for the function $\vphantom{}_0 F _5$.

We then consider that the solution of the problem is a linear combination of the functions of Eqs. \ref{eq:solutions}, over which we have to apply boundary conditions. First of all, boundary condition at infinity rules out  functions $U_j(s)$ with odd index, since these functions diverge when $z\rightarrow \infty$, i.e. when $s\rightarrow 0$. At the interface the first order of fluid velocity should verify
\begin{eqnarray}
v_{1,z}&=&0 \label{eq:bc1}\\
\frac{dv_{1,z}}{dz}&=&0 \label{eq:bc2}
\end{eqnarray}
Conditions for the concentration at the interface can be worked out to obtain
\begin{equation}
\left(\frac{d}{dz}-\mathcal{S}(t)\right)\mathcal{M}\,v_{1,z}=0, \label{eq:bc3}
\end{equation}
where 
\begin{eqnarray}
\mathcal{S}(t)&=&(K-1)\left(1+\dot{\zeta}(t)\right).\label{eq:def_st}
\end{eqnarray}

Now, from Eqs. \ref{eq:bc1}, \ref{eq:bc2}, \ref{eq:bc3} applied over a linear
combination of the three solution functions well-behaved at infinity, i.e. those with even index, we obtain a
homogeneous system of three equations with three unknowns.
By writing
\begin{equation}
v_{1,z}(s)=a_1 U_0(s)+a_2 U_2(s)+a_3 U_4(s),
\end{equation}
this system can be written as a $3 \times 3$ matrix $\mathrm{C}$ such that:
\begin{equation}
\mathrm{C}_{1j}=U_{2(j-1)},\;\;\mathrm{C}_{2j}=\delta\:U_{2(j-1)},
\;\;\mathrm{C}_{3j}=\left(\delta+l\,\mathcal{S}(t)\right)\tilde{\mathcal{M}}\,U_{2(j-1)}
\end{equation}
where we have used the notation $\delta=s\: d\!/\!ds$ and $\tilde{\mathcal{M}}$
for the operator $\mathcal{M}$ once we changed the variables from $\xi$ to
$s$. In order to find a
non-trivial solution, $Det(\mathrm{C})$ has to be equal to zero. The expression of 
$Det(\mathrm{C})$ can then be worked out through a long calculation by exploiting several properties of the generalized hypergeometrical functions, with the result:
\begin{equation}
Det(\mathrm{C})\,s^{-3+r_0+r_2+r_4}=\left(1-r_1+l(t)\,\mathcal{S}(t)\right)Det(\mathrm{A})+Det(\mathrm{B})\label{eq:disprelABC}
\end{equation}
where the elements of the matrices $\mathrm{A}$ and $\mathrm{B}$ are given in the appendix \ref{matrices-ab}. The dispersion relation takes then the form
\begin{equation}
\left(1-r_1+l(t)\,\mathcal{S}(t)\right)Det(\mathrm{A})+Det(\mathrm{B}) = 0\label{eq:disprelAB}
\end{equation}

\subsection{Asymptotic expansion for large $Ra_S$}

The dispersion relation as derived above is of limited utility. In principle Eq. \ref{eq:disprelAB} can be solved numerically. In fact this can be done only with difficulty, and only for not too high values  of Rayleigh number. This was already noticed by Hurle {\it et al.} \cite{HurleJCG1982} in a related problem, who found a region in the $Ra_S$-$a$ plane that was very hard to explore numerically. They called it {\it unrepresentable singular region}. Furthermore, it is difficult to extract general properties of these solutions, due to the fact that they depend on determinants of complicated hypergeometric functions. For instance, numerically one finds some modes from the multiple solutions of Eq. \ref{eq:disprelAB}, but it is difficult to generalize such result from the analytical properties of this equation.

In order to obtain more useful expressions for these determinants, we perform the limit $s\rightarrow-\infty$. Physically this limit corresponds to large $Ra_S$ and a region not very far from the interface, precisely the region where the solute gradient exists and the instability is expected to occur.
The asymptotic expansion of the generalized hypergeometrical functions in this limit is outlined in Appendix \ref{asymptotics}. A cumbersome calculation provides the dispersion relation of the instability as the solution of the following transcendental equation:
\begin{equation}
\left[1-r_1+N_{3,1}+l\mathcal{S}(t)\right]\left(\sqrt{3}+
\mathrm{tan}\left(6s^{\frac{1}{6}}+
3\pi \gamma _0+\pi\left(r_{0}+r_{2}+r_{4}\right)\right)\right)+s^{\frac{1}{6}}=0
\label{eq:disprel_asym}\end{equation}
where $N_{3,1}$ has the following value:
\begin{eqnarray}
N_{3,1}&=&\frac{l}{6} \left(2-Sc^{-1}\right) \left(\dot{\zeta}+1\right)-
\frac{l^2}{12} \left(Sc^{-1} \left(5\, Sc^{-1}-2\right)+5\right) 
\left(\dot{\zeta}+1\right)^2
\label{eq:def_N31}\\
&-& l^2
 \left(Sc^{-1}+1\right)\, \sigma +
\frac{l}{2} \sqrt{4 a^2+\left(\dot{\zeta}+1\right)^2+4\, \sigma }
-3\, l^2 a^2-\frac{25}{144}\nonumber
\end{eqnarray}

From Eq. \ref{eq:disprel_asym} we discern directly several features. On the one hand, this is a
transcendental equation, but much simpler than Eq. \ref{eq:disprelAB}, and then insight on their solutions can be obtained by intuitively solving it for instance graphically. On the other hand the tangent is a periodic function. This implies directly that there will be an infinitude of solutions for this equation as $Ra_S/Sc$ is increased, a feature that was not immediate from the closed form of A and B and a
property that is confirmed numerically after much difficulty for several modes. These solutions are bifurcations of the unstable ``flat'' state, and they could have a role in the full non-linear dynamics of the system.

Finally, the asymptotic expansion provides an additional result. 
We have proven in Appendix \ref{asymptotics} that if we use only the dominant terms in the expansion, the determinants cancel out exactly, and terms beyond all orders must be added to the expansion in
order to obtain Eq. \ref{eq:disprel_asym}. This arises immediately the question of the numerical
precision of the direct computation of the determinants of Eq. \ref{eq:disprelAB}, since the larger part of the
numerical value of each matrix element will cancel out exactly. We ascertain that this is
in fact the origin of the unrepresentable singular region of Hurle {\it et al.}\cite{HurleJCG1982}

\section{Numerical Results and Discussion}
We can now solve  Eq. \ref{eq:disprelAB} and find the stability boundary as a function of time, making $Re(\sigma)=0$ (no oscillating instabilities have been found). The control parameters will be $Ra_S$ and $\tau$.

If we fix $\tau$ and change $Ra_S$ we expect that, as $Ra_S$ rises, the
instability will be anticipated, since the steady system is more unstable with
higher values of $Ra_S$. Nevertheless, little can be said a priori about the first
unstable wavelength.

\begin{figure}
\includegraphics[width=0.75\textwidth]{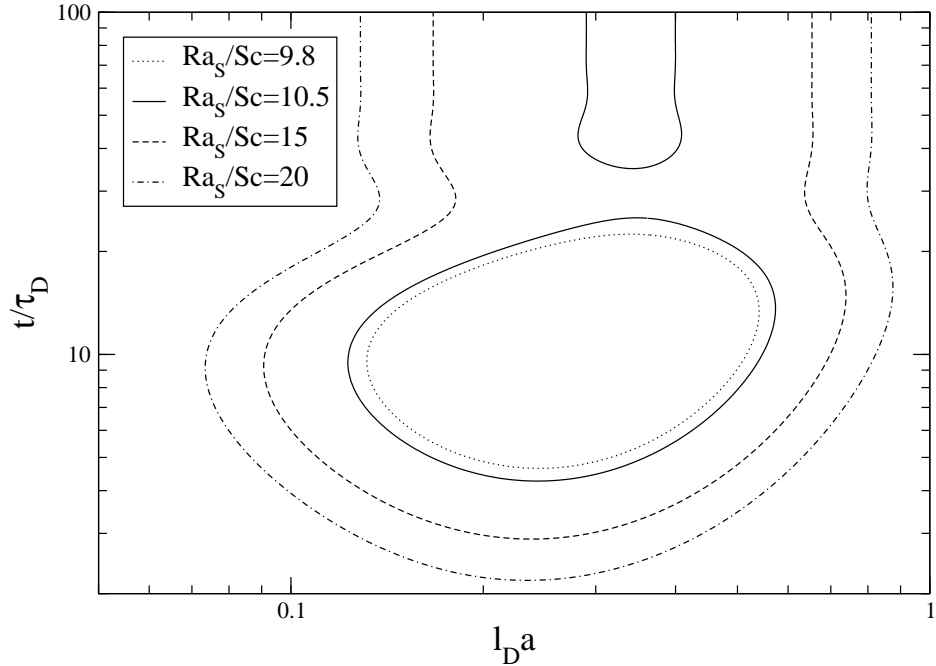}
\caption{Stability boundaries as a function of time for $Sc=81$, $\tau=0.8$, $K=0.3$, 
and several values of the Rayleigh number. The dotted line is the boundary 
 of a transiently unstable region, since its corresponding Rayleigh number 
($Ra_S/Sc=9.8$) is
  slightly below the threshold of the instability in the steady state ($Ra_S/Sc=10.28$).} 
\label{fig:trans_stab_conv_taufix}
\end{figure}

From Fig. \ref{fig:trans_stab_conv_taufix} we see a dramatic dependency
of the stability balloon on $Ra_S$. For values slightly below the threshold, we
see that a transient instability is present, and as we rise the Rayleigh
number above the threshold this region remains detached from the steady
instability until it eventually connects with it.

This behavior can be easily explained if we take into account the actual form
of the stability parameter. From the changes of variables leading to Eq.
\ref{eq:linear_for_v1} we see that the argument of the Hypergeometric function
will be indeed $Ra_S/Sc$, but supplemented with time-dependent factors:

\begin{equation}
-\frac{l^5(t) a^2 Ra_S}{Sc} \left(1-(1-\tau)\zeta(t) \right)
\end{equation}

In particular, we see that this combination of parameters depends on $l(t)$ to
the fifth power, hence its effect will be very important. It is
known\cite{WarrenPRE1993} that values of $\tau$ close to $1$
(equivalently high values of $l_T/l_D$, for instance for large pulling velocities) imply that $l$, when approaching
its steady state value, overshoots severely. This overshooting is therefore
responsible for the transient instability as well as the intricate shapes of
the stability balloon in Fig. \ref{fig:trans_stab_conv_taufix}.
On the other hand, the value of the first unstable wavevector does not change
in an appreciable way when we merely rise the Rayleigh number.

\begin{figure}
\includegraphics[width=0.75\textwidth]{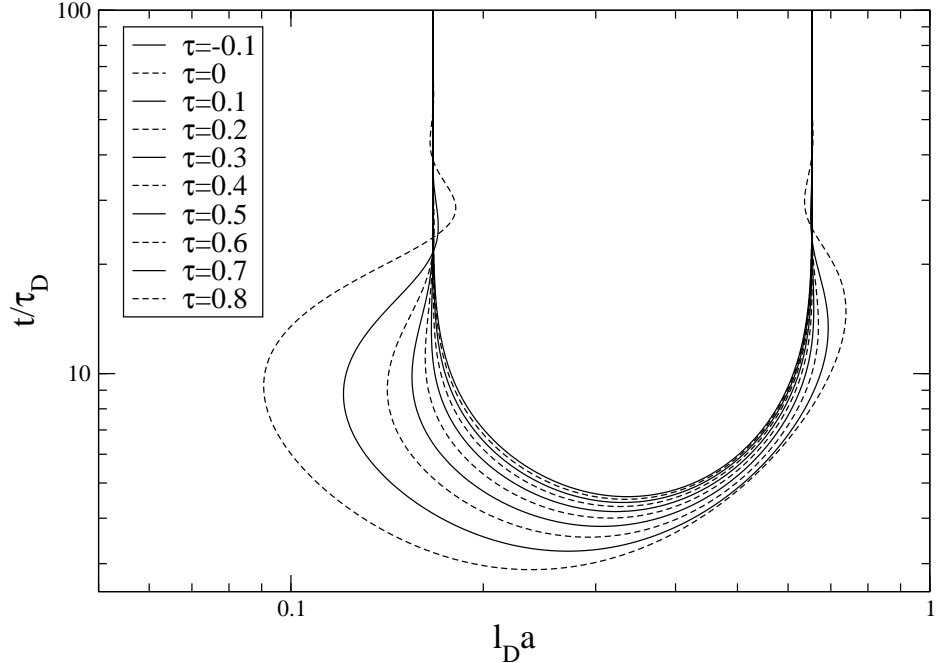}
\caption{Stability boundaries as a function of time for $Ra_S/Sc=15$, $Sc=81$, $K=0.3$, 
and several values of $\tau$. The innermost curve is the one with $\tau=-0.1$,
and the values of the $\tau$ parameter grow outwards. }
\label{fig:trans_stab_conv_rafix}
\end{figure}

We can also consider the effect of $\tau$. From the previous discussion we
also expect $\tau$ to be destabilizing, since values closer to one will mean
higher overshooting. This is the case, as we can see in Fig.
\ref{fig:trans_stab_conv_rafix}. We also see in this figure that, despite the
first unstable wavelength seemed to be almost independent of the Rayleigh
number, higher values of $\tau$ push this wavevector to higher
wavelengths.

This can be understood from the values of the parameters of the hypergeometric
functions (\ref{eq:def_rtilde}). Whenever $a$ appears in the equations, it has
always an $l$ multiplying it, although not necessarily directly. Therefore,
higher $l$ implies smaller $a$.

In principle, it might seem that a natural assumption would be to rescale all lengths with $l$ and assume that the steady states results are still valid in the transient. This was indeed the path followed by Jamgotchian et al.\cite{jamgotchian01}. The Rayleigh number is rescaled by $l^3$, and $\sigma$ by $l^2$, in accordance with the construction of the diffusion time from the diffusion length.

We argue that this approach can only work partially. A simple calculation shows that when $\zeta=0$ the scaling is indeed perfect, but only in that case. For this value of $\zeta$, $(1+\dot{\zeta})=l^{-1}$, and hence the $r_i$ in equation \ref{eq:def_rsintilde} are identical to their counterpart in the steady state. Similarly, the argument of the hypergeometric function or the product $l(t)\,\mathcal{S}(t)$ becomes independent of time, thus reverting to the steady-state case.

\begin{figure}
\begin{center}    
\leavevmode
\includegraphics[height=0.6\textwidth]{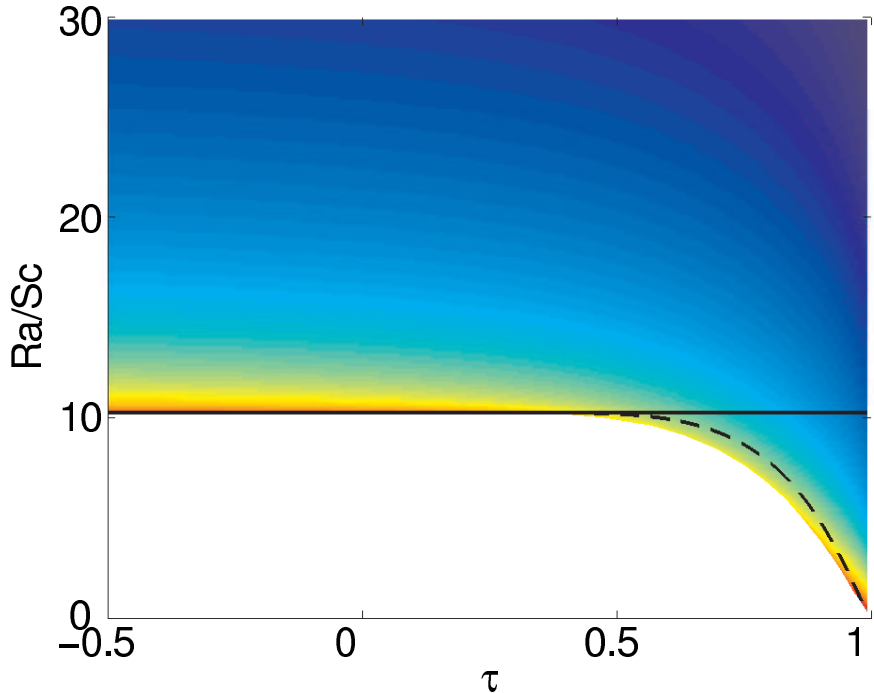}
\end{center}       
\caption{Stability map for $Sc=81$, $K=0.3$. Color depends on
  the logarithm of the time to the first instability (blue -- short, red --
  long). The solid line is the steady-state stability boundary of the convective instability, the dashed line
  corresponds to the instability when performing the scaling of all lengths with $l$.  }
\label{fig:fulltransient}                  
\end{figure}

Nevertheless, even if the transient instability is controlled solely by the value of $l$, the instability will develop in general before the time at which $\zeta = 0$. In Fig. \ref{fig:fulltransient} it is depicted the comparison of the actual time to instability, the boundary of the steady instability and the boundary of the transient instability as computed with the scaling approximation for $K=0.3$ and $Sc=81$.

To begin with, we observe that for small or negative values of $\tau$ the steady instability line is almost coincident with the boundary of the transient instability, and close to the threshold it takes longer for the instability to develop, as it could be expected. For larger values of $\tau$ we see that the boundary of the transient instability goes below the critical $Ra_S$ of the steady instability, i.e. there appears a purely transient instability, as shown in Fig. \ref{fig:trans_stab_conv_taufix}. We see that the scaling approximation (dashed line) is a very good one in that, despite its simplicity, correctly predicts the dependency of $Ra_S$ with $\tau$ for the instability boundary. Nevertheless, we see that the transient instability takes place for smaller values of $Ra_S$, as expected from the previous argument that the scaling is correct for $\zeta=0$ but only approximate when this is not the case.

\begin{figure}
\begin{center}    
\leavevmode
\includegraphics[width=0.75\textwidth]{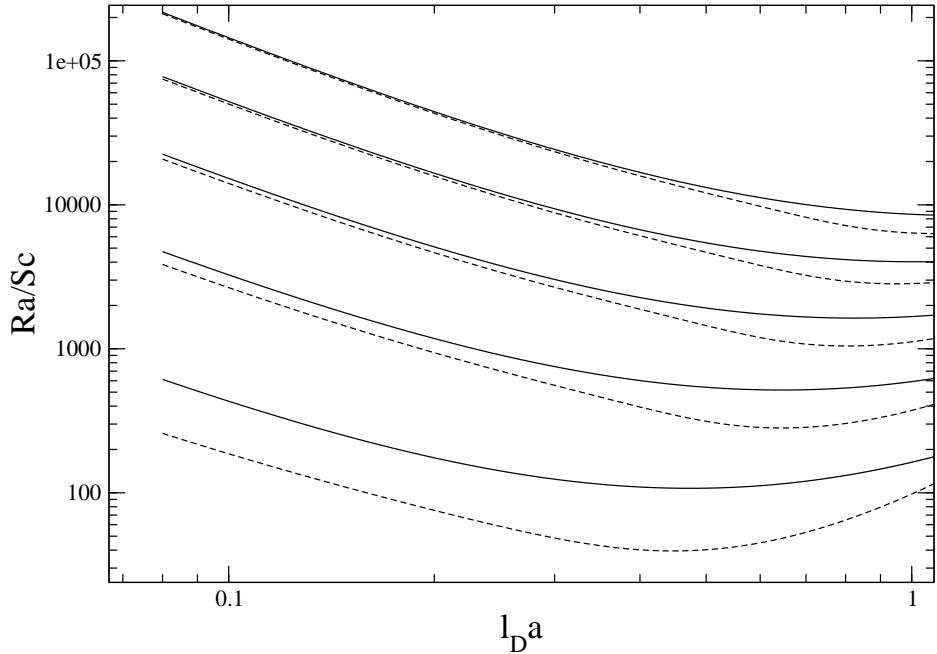}
\leavevmode
\end{center}       
\caption{Successive instabilities of the homogeneous state (modes 2, 3, 4, 5
  and 6) for $Sc=81$, $K=0.3$. Solid: Numerical solution. Dashed: Asymptotic approximation.}
\label{fig:asymptoticfit}                  
\end{figure} 

We can compare now the asymptotic approximation with the numerical results in the low $a$ regime for some of the
lowest-lying modes of the instability in the steady state (see Fig. \ref{fig:asymptoticfit}). We
see that the asymptotic approximation correctly predicts the $a^{-2}$
dependency for small $a$, and we also see that the accuracy of the
approximation increases with $Ra_S$. Nevertheless, we see that for larger values
of $a$ the approximation is not so good. In fact, the asymptotic behavior for
large $a$ is not correctly reproduced. We believe that this is due to the fact
that the arguments of the hypergeometric functions, which in turn depend on
the $r_i$, are the same in the large $a$ limit. This corresponds to a
degenerate problem, and our method cannot be applied in such circumstances.

Finally, the consistency of the approximations performed should be tested against the numerical results. In particular, it should be checked that the adiabatic or quasi-static approximation is indeed a good one. Obviously, near the threshold of the instability this is not the case, since at that point the evolution of the perturbation is extremely slow ($\sigma\approx 0$). Therefore, the location of the threshold will not be precisely determined, but the approximation will still be useful to locate regions where perturbations will be either amplified or damped.
 To test the validity of the approximation and, in turn, discuss the observability of the transient instability, we have computed the growth rate $\sigma$ of the perturbation for the most unstable mode in the transient instability found for $Ra_S/Sc=9.8$ and $\tau=0.8$ (see Fig. \ref{fig:trans_stab_conv_taufix}). The value found is $\sigma{}\tau{}_{D}=0.14$ for $t/\tau{}_D=10$. Hence, even in such an extreme situation, the separation of scales is valid ($t\gtrsim 1/\sigma$). In addition, at that point the derivative of $\sigma$ with respect to time is close to zero, being as it is close to a maximum. Therefore, the time evolution of the flat interface is slow enough for the adiabatic approximation to give consistent results and for the transient instability to develop. 

\section{Conclusions}

We have studied the convective stability of the melt during the initial transient of a directional solidification experiment in a vertical configuration. The setup is such that thermal gradient is stabilizing but the solute layer built during the transient is destabilizing.
 
By using a quasi-stationary approximation we have obtained the time-dependent dispersion relation for perturbations (both in fluid velocity and in concentration) acting over the quiescent solution in which the solute field follows the Warren-Langer approximation for the transient of a planar solidification front. Such dispersion relation has a very complicated analytical structure and appears as very hard to evaluate numerically. We have further performed an asymptotic expansion for large Rayleigh numbers which has permitted on the one hand to extract some properties of the solution and on the other hand to explain the origin of the difficulties that affected the numerical calculations. In particular the asymptotic procedure reveals that the dominant terms of the dispersion relation cancel out exactly. The result comes from sub-dominant terms, and hence is much harder to evaluate numerically from the complete solution. The analytical form of the asymptotic solution also shows the existence of an infinite number of bifurcations of the unstable state.

The numerical evaluation of the dispersion relation has permitted to work out the stability boundary on the parameter space as a function of time. As it could be expected large values of $\tau$ (i.e. small thermal gradients) tend to destabilize the flow, and to displace the instability to larger wavelengths. In this regime, increasing Rayleigh number to above the threshold  a transient instability appears, i.e. the quiescent solution is stable for the steady (large times) state, but it has not been so during the transient. This instability could deform the solidification front and drive the system to a very different final state, which would not have been revealed in a purely steady state analysis. By increasing $Ra_S$ an instability at large times appears, and by increasing it further both instability regimes merge. This could be relevant for microstructure predictions in solidification processes.

In this study we have not considered the possible deformation of the interface. In fact the solidification front does eventually undergo a morphological instability, which in the directional solidification setup occurs for velocities above a threshold value.\cite{Caroli_Book} In this instability the transient considered here turns out to be of capital importance.\cite{CaroliJCG1993} An interesting continuation of this work would be the complete analysis of the coupling between the morphological and the convective dynamics (already analyzed for the steady state\cite{CaroliJPhys1985,CoriellJCG1980}) during the transient. Such analysis would permit to discern the regimes in which both dynamics effectively couple to each other in the initial transient, presumably giving rise to different results from what given by each of them separately.

\begin{acknowledgments}
We acknowledge financial support from Ministerio de Ciencia e Innovaci\'on (Spain) through Project FIS2009-13360-C03-03.
\end{acknowledgments}

\appendix

\section{Explicit expressions for matrices $\mathrm{A}$ and $\mathrm{B}$}
\label{matrices-ab}
The elements of the matrices $\mathrm{A}$ and $\mathrm{B}$ can be
written as:
\begin{eqnarray}
\mathrm{A}_{11}\;=&\mathrm{B}_{11}&=
 \vphantom{}_0F_5(;1+r_1-r_0,1+r_2-r_0,1+r_3-r_0,1+r_4-r_0,1+r_5-r_0;s)\nonumber\\
\mathrm{A}_{12}\;=&\mathrm{B}_{12}&=
 \vphantom{}_0F_5(;1+r_0-r_2,1+r_1-r_2,1+r_3-r_2,1+r_4-r_2,1+r_5-r_2;s)\nonumber\\
\mathrm{A}_{13}\;=&\mathrm{B}_{13}&=
 \vphantom{}_0F_5(;1+r_0-r_4,1+r_1-r_4,1+r_2-r_4,1+r_3-r_4,1+r_5-r_4;s)\nonumber
\end{eqnarray}
\begin{eqnarray}
\mathrm{A}_{21}&=&\mathrm{B}_{21}\;=
(r_1-r_0)\vphantom{}_0F_5(;r_1-r_0,1+r_2-r_0,1+r_3-r_0,1+r_4-r_0,1+r_5-r_0;s)\nonumber\\
\mathrm{A}_{22}&=&\mathrm{B}_{22}\;=
(r_1-r_2)\vphantom{}_0F_5(;1+r_0-r_2,r_1-r_2,1+r_3-r_2,1+r_4-r_2,1+r_5-r_2;s)\nonumber\\
\mathrm{A}_{23}&=&\mathrm{B}_{23}\;=
(r_1-r_4)\vphantom{}_0F_5(;1+r_0-r_4,r_1-r_4,1+r_2-r_4,1+r_3-r_4,1+r_5-r_4;s)\nonumber
\end{eqnarray}
\begin{eqnarray}
&\mathrm{A}_{31}\;=&
\prod_{i=2}^{5}(r_i-r_0)\vphantom{}_0F_5(;1+r_1-r_0,r_2-r_0,r_3-r_0,r_4-r_0,r_5-r_0;s)\nonumber\\
&\mathrm{A}_{23}\;=&
\frac{s\,\vphantom{}_0F_5(;2+r_0-r_2,2+r_1-r_2,1+r_3-r_2,1+r_4-r_2,1+r_5-r_2;s)}{(1+r_0-r_2)(1+r_1-r_2)}\nonumber\\
&\mathrm{A}_{33}\;=&
\frac{s\,\vphantom{}_0F_5(;2+r_0-r_4,2+r_1-r_4,1+r_2-r_4,1+r_3-r_4,1+r_5-r_4;s)}{(1+r_0-r_4)(1+r_1-r_4)}\nonumber
\end{eqnarray}
\begin{eqnarray}
&\mathrm{B}_{31}\;=&
\prod_{i=1}^{5}(r_i-r_0)\vphantom{}_0F_5(;r_1-r_0,r_2-r_0,r_3-r_0,r_4-r_0,r_5-r_0;s)\nonumber\\
&\mathrm{B}_{33}\;=&
\frac{s\,\vphantom{}_0F_5(;2+r_0-r_2,1+r_1-r_2,1+r_3-r_2,1+r_4-r_2,1+r_5-r_2;s)}{(1+r_0-r_2)}\nonumber\\
&\mathrm{B}_{33}\;=&
\frac{s\,\vphantom{}_0F_5(;2+r_0-r_4,1+r_1-r_4,1+r_2-r_4,1+r_3-r_4,1+r_5-r_4;s)}{(1+r_0-r_4)}\nonumber
\end{eqnarray}
 Note that although the elements of $\mathrm{A}$ and
$\mathrm{B}$ are defined as functions of $s$ they are to be evaluated at
$z=\zeta (t)$


\section{Asymptotic expansion of the dispersion relation}
\label{asymptotics}

Conventional asymptotics for the $_p F _q$ function ($p \neq 0$) for large arguments makes use of the Mellin-Barnes representation. By shifting the integration contour, the sum of the residues on the poles of $\Gamma(\alpha _j +s)$ gives an asymptotic expansion for large argument\cite{Whittaker+Watson}, ignoring exponential terms beyond all orders.\cite{Luke} When $p=0$, there are not that kind of poles, and hence the expansion is necessarily exponential.\cite{Luke} By working out this expansion, particularized for $q=5$, we can reach the following expression, in which the expansion can be written as the sum of three series:
\begin{eqnarray}
_0F_5(;\rho_q;z)&\sim&\left[\prod_{i=1}^5\Gamma(\rho_i)\right]\, \frac{1}{4\sqrt{3\pi^5}}
\left(S_0 + S_1 + S_2\right),
\end{eqnarray}
with
\begin{eqnarray}
S_{0}(\rho_q;z)&=&z^{\gamma}
e^{3\sqrt{3}z^{1/6}}\sum_{r=0}^{\infty}N_r\mathrm{cos}\left(\pi\gamma-\frac{\pi
    r}{6}+3z^{1/6}\right)z^{-\frac{r}{6}},\label{eq:s0}\\
S_{1}(\rho_q;z)&=&-\;z^{\gamma}\sum_{r=0}^{\infty}\sum_{j=1}^5
N_r\mathrm{cos}\left(3\pi\gamma-\frac{\pi
    r}{2}+6z^{1/6}+2\pi\rho _j\right)z^{-\frac{r}{6}},\label{eq:s1}\\
S_{2}(\rho_q;z)&=&z^{\gamma}e^{-3\sqrt{3}z^{1/6}}\sum_{r=0}^{\infty}\sum_{j<k}^5
N_r\mathrm{cos}\left(5\pi\gamma-\frac{5\pi
    r}{6}+3z^{1/6}+2\pi(\rho _j+\rho _k)\right)z^{-\frac{r}{6}}.\label{eq:s2}
\end{eqnarray}
In these series, the coefficients $N_k$ fulfill the following
recursion formula
\begin{eqnarray}
k\,N_k=\sum _{s=1} ^5 T_{5-s}(s-k) 6 ^{-s-1} N_{k-s}\,,\;\;\; k &>& 0\\
 N_0 = 1, \;\;\; N_k = 0,\;\;\;k&<&  0 ;
\label{eq:recur-nk}
\end{eqnarray}
where
\begin{eqnarray}
T_s(-k)&=&\sum _{r=0} ^s \frac{(-)^{s-r}T(r-k)}{r!(s-r)!},\label{eq:ts}\\
T(t)&=&\prod_{j=0}^5 \left(t+\omega _j\right),\label{eq:tt}\\
\omega _j &=& 6 (\rho _j -1+ \gamma)\label{eq:omega}.
\end{eqnarray}
We follow the notation in which $\rho_0 = 1$, and the parameter $\gamma$ is given by
\begin{eqnarray}
\gamma &=& \frac{5}{12}-\frac{1}{6}\sum_{i=1}^5\rho _i.\label{gamma}
\end{eqnarray}

We next use this expansion to find an approximation in the limit $s\rightarrow -\infty$ of the determinants of the matrices $\mathrm{A}$ and $\mathrm{B}$.
To do so, some shorthand notation is needed. We will consider a general matrix whose elements are generalized
hypergeometric functions. If the element $(i,j)$ of one of these matrices is
$_0F_q(;\rho_{ij,1},...,\rho_{ij,5};s)$ then the quantities defined in Eqs. \ref{eq:recur-nk},\ref{eq:ts},\ref{eq:tt},\ref{eq:omega} can be computed for each $(i,j)$, by taking these specific values of $\rho_q$, and hence can be re-defined as matrices.

Now, when introducing these series into the matrix elements of $A$ and $B$ given in Appendix \ref{matrices-ab}, it can be seen that most of the factors accompanying the generalized hypergeometric functions can be absorbed, after some manipulations with the series, into the factor $\prod_{i=1}^5\Gamma(\rho_i)$, which becomes the same for all the elements. There are still a $-z$ factor on some elements of the third row. This can be eliminated by redefining their specific $\gamma _{ij}$ values with the substitution
$
\gamma _{32} \rightarrow \gamma '_{32}+1,
$
$
\gamma _{33} \rightarrow \gamma '_{33}+1,
$ 
where the prime denotes the new value of gamma. With these substitutions a
simple calculation shows that, for matrix $\mathrm{A}$:
\begin{eqnarray}
\gamma _{2j}&=&\gamma _{1j}+\frac{1}{6},\\
\gamma _{3j}&=&\gamma _{1j}+\frac{2}{3}.
\end{eqnarray}
For matrix $\mathrm{B}$:
\begin{eqnarray}
\gamma _{2j}&=&\gamma _{1j}+\frac{1}{6},\\
\gamma _{3j}&=&\gamma _{1j}+\frac{5}{6}.
\end{eqnarray}
The result is that, discarding the global factor $\prod_{i=1}^5\Gamma(\rho_i)$, every element $(i,j)$ of both matrices $A$ and $B$ can be written as a  sum of three contributions $S_{ij,0}$, $S_{ij,1}$, $S_{ij,2}$, each one written as in Eqs. \ref{eq:s0},\ref{eq:s1},\ref{eq:s2}, but with the understanding of taking the element $(i,j)$ of each quantity depending on the $\rho_q$.

The resulting series depend on the  $N_r$ coefficients, which theirselves depend on
the actual parameters of the hypergeometric function through a recurrence
relation which involves a polynomial whose roots are $\omega _j = \beta
(\rho _j -1+ \gamma)$. We are now going to use the actual values of the $\rho _q$ to prove that in
fact for each of the functions the roots are the same up to a permutation.

To do that we can compute the matrix elements of $\omega_k$ for each determinant and each row. First of all, particularizing for the first row of both $A$ and $B$ matrices, it can be seen that the set of the $\{\omega^{1j}_k\}_{k=0, 1 \dots 5}$ are in fact just permutations for different j. 
Therefore, since the roots of the polynomial can be exchanged, the value of
$N^{1j}_r$ { does not depend on} $j$.
For the second row of both $A$ and $B$ we notice that the coefficients of the
hypergeometric functions on that row are in fact the same if we make the
substitution $r_1 \rightarrow r'_1+1$. We can then proceed as in the previous row and prove that 
$N_{2j,r}$ does not depend on $j$ either, and its value is the same. Finally, for the third row, we have to distinguish between
$\mathrm{A}$ and $\mathrm{B}$.
For matrix $\mathrm{A}$, the substitutions
$r_0 \rightarrow r'_0-1$, $r_1 \rightarrow r'_1-1$ will bring us to the first row, and on matrix $\mathrm{B}$, this will be
accomplished by the substitution $r_0 \rightarrow r'_0-1 $. Therefore $N_{3j,r}$ is independent of $j$, and the value is also the same as the other rows. The final conclusion is that the coefficients of the asymptotic expansion $N_r$ do not depend on the particular eigenfunction.

We are now ready for expanding the determinants of the matrices $A$ and $B$.
We introduce the following notation:
\begin{equation}
\mathcal{T}^{rst}_{lmn}
=
\left|
\begin{array}{l}
S_{1j,l}(r)\\ S_{2j,m}(s)\\ S_{3j,n}(t)
\end{array}
\right|
\end{equation}
where $S_{1j,l}(r)$ stands for the whole row, not just the $j^{th}$ element.
Written in that way we can express the whole series, by using the
properties of the determinants, as
\begin{equation}
\left|\mathrm{A}\right| \sim 
\sum _{l}\sum _{m}\sum _{n}\sum _{r}\sum _{s}\sum _{t}  
\mathcal{T}^{rst}_{lmn}
\end{equation}


\subsection{Dominant terms}

With the previous definitions, the dominant terms can be
denoted by $\mathcal{T}^{rst}_{000}$.
The corresponding determinant has the form:
\begin{equation}
\small
K\left|
\begin{array}{ccc}
\mathrm{cos}\left(\pi\gamma _1-\frac{\pi r}{6}+3z^{1/6}\right)&
\mathrm{cos}\left(\pi\gamma _2-\frac{\pi r}{6}+3z^{1/6}\right)&
\mathrm{cos}\left(\pi\gamma _3-\frac{\pi r}{6}+3z^{1/6}\right)\\
\mathrm{cos}\left(\pi\gamma _1-\frac{\pi (s-1)}{6}+3z^{1/6}\right)&
\mathrm{cos}\left(\pi\gamma _2-\frac{\pi (s-1)}{6}+3z^{1/6}\right)&
\mathrm{cos}\left(\pi\gamma _3-\frac{\pi (s-1)}{6}+3z^{1/6}\right)\\
\mathrm{cos}\left(\pi\gamma _1-\frac{\pi (t-4)}{6}+3z^{1/6}\right)&
\mathrm{cos}\left(\pi\gamma _2-\frac{\pi (t-4)}{6}+3z^{1/6}\right)&
\mathrm{cos}\left(\pi\gamma _3-\frac{\pi (t-4)}{6}+3z^{1/6}\right)\\
\end{array}
\right|
\normalsize \nonumber
\end{equation}
where
\begin{equation}
K=N_{1,r} N_{2,s} N_{3,t} z^{\gamma _1 +\gamma _2 + \gamma
_3-(r+s+t-5)/6} e^{9\sqrt{3}z^{1/6}}.
\end{equation}
We have used the definitions of the $\gamma _{ij}$ for $\mathrm{A}$ (this
choice is immaterial, as it will be clearly shown) and the substitutions
$N_{i,r}=N_{ij,r}$, $\gamma _{j}=\gamma _{1j}$.
Expanding the cosine functions of the second and third rows, and
using the properties of the determinants, it is very simple to show that
\begin{equation}
\mathcal{T}^{rst}_{000}=0
\end{equation}
i.e., {\it all the dominant terms in the asymptotic expansion of the
  determinant cancel exactly.}

\subsection{Beyond all orders}

Since all the dominant terms cancel, sub-exponential terms are needed. 
The first sub-exponential terms are of the form:
\begin{equation}
\mathcal{T}^{rst}_{100}
\end{equation}

These terms will be computed in their general form, but before some more notation is required.
We make the following definitions:
\begin{eqnarray}
\gamma{}_0 &=& -\frac{5}{12}-\frac{1}{6}\sum_{i=0}^5 r_i\\
\alpha{}_0 &=& -\frac{r}{6}\\
\alpha{}_1 &=& \frac{1-s}{6}\\
\alpha{}_2 &=& \frac{4-t}{6} \:\mathrm{(matrix\,A)}\\
\alpha{}_2 &=& \frac{5-t}{6} \:\mathrm{(matrix\,B)}\\
K &=& N_{1,r} N_{2,s} N_{3,t} z^{\gamma _1 +\gamma _2 + \gamma
_3-(r+s+t-5)/6} e^{6\sqrt{3}z^{1/6}} \:\mathrm{(matrix\,A)}\\
K &=& N_{1,r} N_{2,s} N_{3,t} z^{\gamma _1 +\gamma _2 + \gamma
_3+1-(r+s+t)/6} e^{6\sqrt{3}z^{1/6}} \:\mathrm{(matrix\,B)}
\end{eqnarray}

With these definitions, the general determinant can be written as:
\begin{eqnarray}
\mathcal{T}^{rst}_{100}&=&-K\epsilon{}_{ijk}\sum_{l=1}^5
\mathrm{cos}\left(6z^\frac{1}{6}+3\pi{}(\gamma{}_0+\alpha{}_0)+\pi r_{2(i-1)}+
2\pi{}r_{l-\theta{}(2(i-1)-l)}\right)\\
& &\mathrm{cos}\left(3z^\frac{1}{6}+\pi{}(\gamma{}_0+\alpha{}_1)+\pi r_{2(j-1)}\right)\mathrm{cos}\left(3z^\frac{1}{6}+\pi{}(\gamma{}_0+\alpha{}_2)+\pi r_{2(k-1)}\right),
\end{eqnarray}
in which summation over $i,j,k$ from 1 to 3 is implied. $\epsilon{}_{ijk}$ is the {\it Levi-Civita} symbol, and $\theta(n)$ is a discrete step function with
$\theta(n)=0$ for $n < 0$ and $\theta(n)=1$ for $n \geq 0$.

After a lengthly manipulation of the trigonometric functions, and by using symmetry properties under exchanging of indexes, one can simplify this last expression to get
\begin{eqnarray}
\mathcal{T}^{rst}_{100}&=&-4 K\mathrm{sin}\left(\pi{}(\alpha{}_1-\alpha{}_2)\right)
\mathrm{sin}\left(\pi \left(r_{2}-r_{4}\right)\right)
\mathrm{sin}\left(\pi \left(r_{4}-r_{0}\right)\right)
\mathrm{sin}\left(\pi \left(r_{0}-r_{2}\right)\right) \nonumber \\
& &\mathrm{cos}\left(6z^{\frac{1}{6}}+3\pi \gamma _0+3\pi \alpha _0 + \pi\left(r_{0}+r_{2}+r_{4}\right)\right).
\end{eqnarray}

Notice that in fact we have only computed a separate term. We have assumed a certain position for the sub-exponential term in the determinant, but to compute properly the contribution we should sum for all possibilities:
\begin{displaymath}
\mathcal{T}^{rst}_{100}
+
\mathcal{T}^{rst}_{010}
+
\mathcal{T}^{rst}_{001}.
\end{displaymath}
This is tantamount to a cyclic sum over the $\alpha _{i}$. Further manipulation gives
\begin{eqnarray}
\mathcal{T}^{rst}_{100}+\mathcal{T}^{rst}_{010}+\mathcal{T}^{rst}_{001}&=&
-16 K
\mathrm{sin}\left(\pi{}(\alpha{}_1-\alpha{}_2)\right)
\mathrm{sin}\left(\pi{}(\alpha{}_2-\alpha{}_0)\right)
\mathrm{sin}\left(\pi{}(\alpha{}_0-\alpha{}_1)\right)\nonumber\\
& &\mathrm{sin}\left(\pi \left(r_{2}-r_{4}\right)\right)
\mathrm{sin}\left(\pi \left(r_{4}-r_{0}\right)\right)
\mathrm{sin}\left(\pi \left(r_{0}-r_{2}\right)\right)\label{eq:magicrule}\\
& &\mathrm{cos}\left(6z^{\frac{1}{6}}+3\pi \gamma _0+
\pi\left(\alpha _0+\alpha _1+\alpha _2\right) + 
\pi\left(r_{0}+r_{2}+r_{4}\right)\right)\nonumber
\end{eqnarray}


We can now evaluate the terms corresponding to the first orders in the large-$z$ expansion.
For each matrix A and B we consider first the zeroth order terms:
\begin{displaymath}
\mathcal{T}^{000}_{100}
+
\mathcal{T}^{000}_{010}
+
\mathcal{T}^{000}_{001}.
\end{displaymath}
Note that those terms are not of the same order in $z$ for the matrices A and B, due to the different order of the $K$ for each matrix. Matrix B gives in fact the lower order in $z$.

We start then with the matrix B. For this matrix
the value of the $\alpha _i$ are 
$\alpha=\left[0\;\frac{1}{6}\;\frac{5}{6}\right]$,
which  in Eq. \ref{eq:magicrule} gives:
\begin{eqnarray}
\mathcal{T}^{000}_{100}+\mathcal{T}^{000}_{010}+\mathcal{T}^{000}_{001}&=&
2\sqrt{3}
\mathrm{sin}\left(\pi \left(r_{2}-r_{4}\right)\right)
\mathrm{sin}\left(\pi \left(r_{4}-r_{0}\right)\right)
\mathrm{sin}\left(\pi \left(r_{0}-r_{2}\right)\right)\nonumber\\
& &\mathrm{cos}\left(6z^{\frac{1}{6}}+3\pi \gamma _0+
\pi\left(r_{0}+r_{2}+r_{4}\right)\right)
z^{\gamma _1 +\gamma _2 + \gamma
_3+1} e^{6\sqrt{3}z^{1/6}}
\end{eqnarray}

Order zero of matrix A contributes to the next order in $z$.
For this matrix the value of the $\alpha _i$ are given by
$\alpha=\left[0\;\frac{1}{6}\;\frac{2}{3}\right]$.
If we put that in Eq. \ref{eq:magicrule} we obtain:
\begin{eqnarray}
\mathcal{T}^{000}_{100}+\mathcal{T}^{000}_{010}+\mathcal{T}^{000}_{001}=
-4\sqrt{3}
\mathrm{sin}\left(\pi \left(r_{2}-r_{4}\right)\right)
\mathrm{sin}\left(\pi \left(r_{4}-r_{0}\right)\right)
\mathrm{sin}\left(\pi \left(r_{0}-r_{2}\right)\right)\nonumber\\
\mathrm{cos}\left(6z^{\frac{1}{6}}+3\pi \gamma _0+
\frac{5\pi}{6}+\pi\left(r_{0}+r_{2}+r_{4}\right)\right)
z^{\gamma _1 +\gamma _2 + \gamma
_3+5/6} e^{6\sqrt{3}z^{1/6}}
\end{eqnarray}



We only consider now the next order of matrix B, which is of the same order in $z$ than the zeroth order of matrix A. There are nine of those order 1 terms:
\begin{displaymath}
\mathcal{T}^{100}_{100}+\mathcal{T}^{100}_{010}+\mathcal{T}^{100}_{001}+
\mathcal{T}^{010}_{100}+\mathcal{T}^{010}_{010}+\mathcal{T}^{010}_{001}+
\mathcal{T}^{001}_{100}+\mathcal{T}^{001}_{010}+\mathcal{T}^{001}_{001}
\end{displaymath}
We can evaluate the previous expression by summing terms in groups of 3. For $\mathcal{T}^{100}_{100}+\mathcal{T}^{100}_{010}+\mathcal{T}^{100}_{001}$ the vector of the $\alpha _i$ is $\alpha=\left[-\frac{1}{6}\;\frac{1}{6}\;\frac{5}{6}\right]$.
Since $\alpha _2-\alpha _0 =1$ one of the sines in Eq. \ref{eq:magicrule} will be zero and this term will not contribute. The next group is $\mathcal{T}^{010}_{100}+\mathcal{T}^{010}_{010}+\mathcal{T}^{010}_{001}$.
Here, the vector of the $\alpha _i$ is $\alpha=\left[0\;0\;\frac{5}{6}\right]$. For this case $\alpha _0-\alpha _1 =0$ and again one of the sinus in Eq. \ref{eq:magicrule} will be zero and this term will neither contribute.
For the group of terms $\mathcal{T}^{001}_{100}+\mathcal{T}^{001}_{010}+\mathcal{T}^{001}_{001}$
the vector of the $\alpha _i$ is
$\alpha=\left[0\;\frac{1}{6}\;\frac{2}{3}\right]$.
But this is the same vector that was previously found for $\mathrm{A}$ at order 0. The final result will then be, for the matrix B:
\begin{eqnarray}
\mathcal{T}^{001}_{100}+\mathcal{T}^{001}_{010}+\mathcal{T}^{001}_{001}&=&
-4\sqrt{3}N_{3,1}
\mathrm{sin}\left(\pi \left(r_{2}-r_{4}\right)\right)
\mathrm{sin}\left(\pi \left(r_{4}-r_{0}\right)\right)
\mathrm{sin}\left(\pi \left(r_{0}-r_{2}\right)\right)\nonumber\\
& &\mathrm{cos}\left(6z^{\frac{1}{6}}+3\pi \gamma _0+
\frac{5\pi}{6}+\pi\left(r_{0}+r_{2}+r_{4}\right)\right)
z^{\gamma _1 +\gamma _2 + \gamma
_3+5/6} e^{6\sqrt{3}z^{1/6}}\nonumber
\end{eqnarray}
with $N_{3,1}$ as in Eq. \ref{eq:def_N31}.


Finally, we have all the elements at our disposal to build an approximation to the dispersion relation in the form given by Eq. \ref{eq:disprelABC}. If we substitute our approximation in this formula we obtain:
\begin{eqnarray}
-4\sqrt{3}\left[1-r_1+N_{3,1}+l\mathcal{S}(t)\right]\mathrm{cos}\left(6z^{\frac{1}{6}}
\vphantom{\frac{1}{6}}\right.\!\!\!&\!\!\!+\!\!\!&\!\!\!\left.3\pi \gamma _0+
\frac{5\pi}{6}+\pi\left(r_{0}+r_{2}+r_{4}\right)\right)\\
&+&z^{\frac{1}{6}}2\sqrt{3}\mathrm{cos}\left(6z^{\frac{1}{6}}+3\pi \gamma _0+
\pi\left(r_{0}+r_{2}+r_{4}\right)\right)\;=\;0 \nonumber
\end{eqnarray}
After rearranging the terms we can write it in a more compact form:
\begin{equation}
\left[1-r_1+N_{3,1}+l\mathcal{S}(t)\right]\left(\sqrt{3}+
\mathrm{tan}\left(6z^{\frac{1}{6}}+
3\pi \gamma _0+\pi\left(r_{0}+r_{2}+r_{4}\right)\right)\right)+z^{\frac{1}{6}}=0
\label{eq:disprel_asym-app}\end{equation}
Which, clearly, is a transcendental equation for $z$.
The direct analysis of equation \ref{eq:disprel_asym} shows that there exists an infinite number of solution branches as $z$ increases, due to the periodicity of the tangent.

\bibliography{meca-sol-conv_subted}

\end{document}